\documentclass[preprint,12pt]{elsarticle}

\usepackage{hyperref}
\usepackage{graphicx}
\usepackage{dcolumn}
\usepackage{bm}

\usepackage[utf8]{inputenc}
\usepackage[T1]{fontenc}
\usepackage{mathptmx}
\usepackage{amsmath}
\usepackage{dirtree}
\usepackage{multirow}
\usepackage{subfig}
\usepackage{amssymb}
\usepackage[dvipsnames]{xcolor}

\newcounter{bla}

\journal{Computer Physics Communications}

\begin{document}

\begin{frontmatter}



\title{\texttt{Spinney}: post-processing of first-principles calculations of point defects in semiconductors with Python}


\author[a]{Marco Arrigoni \corref{author}}
\author[a]{Georg K. H. Madsen}

\cortext[author] {Corresponding author.\\\textit{E-mail address:} marco.arrigoni@tuwien.ac.at}
\address[a]{Institute of Materials Chemistry, TU Wien, A-1060 Vienna, Austria}

\begin{abstract}
Understanding and predicting the thermodynamic properties of point defects in semiconductors and insulators would greatly aid in the design of novel materials and allow tuning the properties of existing ones.
As a matter of fact, first-principles calculations based on density functional theory (DFT) and the supercell approach have become a standard tool for the study of point defects in solids. However, in the dilute limit, of most interest for the  design of novel semiconductor materials, the ``raw`` DFT calculations require  an extensive post-processing.
\texttt{Spinney} is an open-source Python package developed with the aim of processing first-principles calculations to obtain several quantities of interest, such as the chemical potential limits that
assure the thermodynamic stability of the defect-laden system, defect charge transition levels, defect formation energies, including electrostatic corrections for finite-size effects, and defect and carrier concentrations.  
In this paper we demonstrate the capabilities of the \texttt{Spinney} code using c-BN, GaN:Mg, TiO$_2$ and ZnO as examples.

\end{abstract}

\begin{keyword}
first-principles; point defects; charged defects; thermodynamic stability; density functional theory; Python

\end{keyword}

\end{frontmatter}



{\bf PROGRAM SUMMARY}

\begin{small}
\noindent
{\em Program Title:}      \texttt{Spinney}                                     \\
{\em Licensing provisions}:  MIT                                  \\
{\em Programming language:}  Python 3                                  \\
{\em Extetrnal libraries:}    NumPy \cite{numpy}, SciPy \cite{scipy}, Pandas \cite{pandas}, Matplotlib \cite{matplotlib}, ASE \cite{ase}                       \\
{\em Nature of the problem:}\\
Post-processing of first-principles calculations in order to obtain important properties of defect laden systems in the dilute-limit: chemical potential values ensuring thermodynamic stability, 
thermodynamic charge transition levels, defect formation energies and corrections thereof using state-of-the-art corrections schemes for electrostatic finite-size effects, equilibrium defect and carriers concentrations.  \\
{\em Solution method:}\\
 Flexible low-level interface for allowing the post-processing of the raw fist-principles data provided by any computer code. High-level interface for parsing and post-processing the first-principles data produced by the popular computer codes VASP and WIEN2k. \\
{\em Additional comments:}\\
An extensive documentation is available at: \href{https://spinney.readthedocs.io}{https://spinney.readthedocs.io} \\
The code is hosted on GitLab: \href{https://gitlab.com/Marrigoni/spinney}{https://gitlab.com/Marrigoni/spinney}
   \\

\end{small}

\section{Introduction}
\label{sec:intro}
Point defects in semiconductor and insulator materials  (from now on we will use the general term ``semiconductor'' to denote any material with a non-zero fundamental gap) play a fundamental role in a wide range of technological applications such as electronics, optoelectronics, electrochemistry and catalysis to name a few \cite{book:Pizzini, book:McCluskey, Queisser-1998, Maier-2005, Alamo-2011, Yu-2016}.
The introduction of controlled amounts of extrinsic atomic species, usually in very low concentrations, can have greatly beneficial effect on  the properties of the host material and has fueled the enormous growth of the semiconductor industry.
At the same time however, the unintentional introduction of impurities can have a detrimental effect on the performance and lifetime of a device. In addition to extrinsic species, intrinsic point defects, whose existence is guaranteed at thermodynamic equilibrium, play a fundamental role in defining the properties of any material and their response to doping \cite{Laks-1991, Zunger-2003, Walsh-2017}.

The desire to control materials properties through point defects has lead to the need of a detailed understanding of the atomic-scale structure and electronic properties of such  defects.
 While it is sometime possible to experimentally obtain details of the atomic structure of surface defects \cite{Setvin-2014, Reticcioli-2017}, acquiring such information remains generally challenging. First-principles density functional theory (DFT)\cite{Hohenberg-1964,Kohn-1965} calculations give the possibility of both atomic-scale structural characterization as well as the calculation of thermodynamic quantities\cite{Walle-2004,DefectsSemiconductors,PDRev}, and have become a standard tool for complementing experimental investigations. 

The vast majority of such calculations employ the supercell approach, in which a point defect is embedded in a simulation cell containing multiple repetitions of  the primitive cell of the host material (\emph{i.e.} the supercell). However, for many practical applications the dilute limit, in which defect-defect interactions can be considered negligible, is of interest. The modeling of such a limit poses a challenge for the supercell approach, as commonly used supercell sizes entail much larger defect concentrations. Moreover, point defects in semiconductors can be ionized and therefore realize different charge states \cite{Leslie-1985}. While most of the spurious defect-defect interactions present in the supercell method can be minimized by employing a moderate supercell size, the long-ranged nature of the Coulomb potential means that correcting spurious electrostatic interactions is fundamental in order to obtain reliable defect formation energies \cite{Makov-1995}. 

The computer code \texttt{Spinney} was developed with the aim of obtaining the most important properties of the defect-laden systems from the results of DFT supercell calculations by implementing non-trivial post-processing operations. While other codes aimed at similar properties support only  calculations performed by specific DFT packages,\cite{Pean-2017, Goyal-2017, Broberg-2018} \texttt{Spinney} offers a flexible low-level interface operating on built-in Python or NumPy data structures. This allows users of general DFT softwares to extract the required information, feed it to the relevant \texttt{Spinney} subroutines and obtain the defect properties of interest without being bound to use code-specific formats and data structures. At the same time, \texttt{Spinney} provides user-friendly interfaces for selected popular DFT codes, able to automatically parse the DFT calculations outputs, extract the needed information and produce properties of interest. At the current stage, such interfaces exist for the VASP \cite{Kresse-1993, Kresse-1996} and WIEN2k \cite{wien2k,blaha_JCP20} first-principles codes.   

In section \ref{sec:background} we review the theoretical background regarding the point defects in the dilute limit and describe the formalism which is employed by \texttt{Spinney}.  Section \ref{sec:examples} gives a demonstration on how \texttt{Spinney} can be used to calculate common properties of defect-laden system using important semiconductor and oxide materials as explicit examples. 


\section{Background}
\label{sec:background}
\subsection{\label{subsec:dilute} Defect formation energy and thermodynamic charge transition levels in the dilute limit}
A key quantity that characterizes a point defect is its formation energy. The most appropriate thermodynamic potential for the formation of point defects is the grand potential. Therefore it is natural to define the formation energy, $\Delta E_d(D^{(q)})$, of a given point defect $D$ in the charge state $q$ as the change in grand potential after  the introduction of the defect in the pristine host material \cite{Zhang-1991}:

\begin{equation}
\label{eq:formene}
\Delta E_d(D^{(q)}) = \Delta E_f(D^{(q)}) - \sum_i n_i \Delta\mu_i + q \mu_e + E_{corr}
\end{equation}

where $\Delta E_f(D^{(q)})=E(D^{(q)})-E_\mathrm{bulk}-\sum_i n_i E_i$ is the defect formation
energy with respect to the reference states of the parent elements. $E(D^{(q)})$ is the energy of the supercell containing the point defect, $E_\mathrm{bulk}$ is the energy of the supercell describing the pristine material, $n_i$ is the number of atoms of type $i$ which need to be removed ($n_i < 0$) or added ($n_i > 0$) to the system in order to create the point defect, $\Delta\mu_i$ is the chemical potential of the $i$-th element with respect to the standard state of the element and $\mu_e$ is the chemical potential of the electron.

As a common approximation,  the Gibbs free energy of the solid has been replaced by the ground-state DFT energy, $E$. For defects inducing large atomic distortions on the host material, harmonic \cite{Arrigoni-2015,Bjor-2016,Arrigoni-2016}, and even anharmonic contributions \cite{Glensk-2014}, can have a non-negligible effect at high temperatures. However, the calculation of such terms is generally time-consuming and is customarily neglected. On the other hand, large errors arise if thermal contributions are neglected in the chemical potential of gas species. Therefore, one needs to explicitly account for the pressure and temperature dependence of gas species if these appear in equation \eqref{eq:formene}. Furthermore, $E(D^{(q)})$ must contain a correction to the "raw" DFT energy due to the finite-size-errors which arise in the supercell method \cite{PDRev}, which is denoted as $E_{corr}$ in equation \eqref{eq:formene}. More detailed discussions of the chemical potentials and finite size corrections are given in Sections \ref{subsec:thermo} and \ref{subsec:electro}.  

A thermodynamic charge transition level is defined as the value that the electron chemical potential must have in order for two different charge states of a defect, $q$ and $q'$, to have the same defect energy. 
It is customarily to express the electron chemical potential in terms of the valence band maximum of the host material, $\epsilon_V$, and the Fermi level, $E_F$, which varies between zero and the band gap of the material: $\mu_e = \epsilon_V + E_F$. With this convention and equation~\eqref{eq:formene}, we can then write the charge transition levels as
\begin{equation}
\label{eq:translevel}
\epsilon_0(q/q') = \frac{E(D^{(q)}) - E(D^{(q')})}{q' - q}- \epsilon_V + \Delta E_{corr}(q/q'),
\end{equation}

The expression emphasizes the fact that charge transition levels do not depend on the chemical potentials of the elements but they do depend on the valence band maximum and the corrections for finite size effects.

\subsection{\label{subsec:thermo} Determination of equilibrium chemical potential values}
Chemical potentials  quantify the energy cost necessary for exchanging atomic species between the system and a particle reservoir. The values of $\mu_i$ and $\mu_e$ are thus important, not only because they determine the defect formation energy, equation~\eqref{eq:formene}, but also because they  connect  the first-principles calculations with the experimental growth conditions.

Thermodynamic equilibrium constrains the possible values the chemical potentials can assume. Considering, without loss of generality, a binary compound, $\mathrm{M_xN_y}$, the values of $\Delta\mu_M$ and $\Delta\mu_N$ are restricted by the following conditions:

\begin{subequations}
\label{eq:atchempots}
\begin{equation}
\label{subeq:equality}
x\Delta \mu_{M} + y \Delta \mu_N = \Delta h_f(\mathrm{M_xN_y}),
\end{equation}
\begin{equation}
\label{subeq:inequality}
\nu \Delta \mu_{M} + \mu \Delta \mu_N \leq  \Delta h_f(\mathrm{M}_\nu\mathrm{N}_\mu),
\end{equation}
\begin{equation}
\label{subeq:bounds}
\Delta \mu_{M} \leq 0 \quad,\quad 
\Delta \mu_N  \leq 0. 
\end{equation}
\label{eq:equal_bound}
\end{subequations}

$\Delta \mu_{\zeta}$ represents the change in chemical potential from the standard state of species $\zeta$: $\Delta \mu_\zeta = \mu_\zeta - \mu^\circ_\zeta $, and $\Delta h_f$ is the enthalpy of formation (per formula unit) of the compound.
Equation \eqref{subeq:equality} represents the thermodynamic stability condition for $\mathrm{M_xN_y}$ and shows that only one of the elemental chemical potentials is an independent variable. 
Equation~\eqref{subeq:inequality} constrains the value of the chemical potentials so that  $\mathrm{M_xN_y}$ is stable and do not segregate into other phases. Likewise, equation~\eqref{subeq:bounds} demands that segregation of the parent compounds are avoided.

As mentioned above, thermal contributions to the free energies of gas phases must be taken into account. Suppose that in the standard state species N is in the gas phase N$_\gamma$, we can  model the dependence of $\Delta\mu_{N_\gamma}$ from $T$ using a set of parameters fitted to experimental data (Shomate equation) \cite{NIST} and model the dependence on $p$ by using an ideal gas model:
\begin{equation}
\label{eq:chem_o2}
\Delta\mu_{N}(T, p) = \frac{1}{\gamma}\left(h_{N_\gamma}(0, p^\circ) + \Delta g_{N_\gamma} (T) + k_B T \ln\left(\frac{p}{p^\circ}\right)\right)
\end{equation} 
where $p^\circ = $ 1 bar is the standard pressure and $\Delta g_{N_\gamma} (T) =  g_{N_\gamma}(T, p^\circ) - g_{N_\gamma}(0, p^\circ) $. $h_{N_\gamma}(0, p^\circ) $ represents the enthalpy per molecule at standard pressure and 0 K and it is usually approximated by the DFT-calculated electronic energy.  This model is implemented in \texttt{Spinney} for the most common binary gas molecules: O$_2$, N$_2$, H$_2$, Cl$_2$ and F$_2$. The model can also be used for other gases taking the needed parameters for calculating $\Delta g_{N_\gamma}$ from the NIST-JANAF themochemical tables \cite{NIST}.

\subsection{\label{subsec:electro} Corrections for electrostatic finite-size effects}
Due to the periodic boundary conditions, the introduction of a point defect in a supercell entails the presence of periodic images of the defect itself.
Such an artificial array of defects furthermore generally corresponds to high defect concentrations, far from the dilute limit. While most of the resulting defect-defect interactions can be minimized by using a large enough supercell, long range electrostatic interactions due to charged defects cannot be neglected for any realistic supercell size. 
In such a case, spurious defect-defect interactions can considerably affect
the predicted energy of the point defect and must be corrected for.

As a point defect is introduced in the host material, there will be a redistribution of the charge density of the latter. In the ideal case of an isolated point defect, this defect-induced charge density can be described by $\rho_{isol}$.
On the other hand, when a supercell is employed, the application of periodic boundary conditions will yield a different charge density: $\rho_{per}$.
Moreover, for charged point defects, periodic boundary conditions require the introduction of a neutralizing background, usually taken as an homogeneous jellium of density $-\frac{q}{V}$, where $q$ is the defect charge state and $V$ is the supercell volume, in order to ensure convergence of the electrostatic energy \cite{Leslie-1985}.
The defect-induced electrostatic potential can generally be obtained from the induced charge density by solving Poisson equation, $\nabla^2 \phi(\mathbf{r}) = -4\pi \rho(\mathbf{r})$, with the proper boundary conditions. This will yield the potentials $\phi_{isol}$ and $\phi_{per}$ for  $\rho_{isol}$ and $\rho_{per}$, respectively.

Assuming that the defect-induced charge density is completely contained in the supercell and is not affected by the presence of periodic boundary conditions, the corrective term can be calculated as:
\begin{equation}
\label{eq:corre}
E_{corr} = \frac{1}{2} \int_V \left(\phi_{isol}(\mathbf{r})-\phi_{per}(\mathbf{r})\right)  \rho_d(\mathbf{r}) \, d\mathbf{r}
\end{equation}
Multiple correction schemes for electrostatic finite-size effects have been developed \cite{Makov-1995, Schultz-1999, Schultz-2000, Zunger-2008, Dabo-2008, Freysoldt-2009, Kumagai-2014, Durrant-2018} in order to estimate $E_{corr}$. 
Among the proposed approaches, the methods applying corrections \emph{a posteriori} using simple models for the defect-induced charge density have become perhaps the most popular due to their flexibility, speed and reliability. 

\texttt{Spinney} implements the \emph{a posteriori} schemes proposed by Freysoldt, Neugebauer and Van der Walle (FNV) \cite{Freysoldt-2009} and the development thereof proposed by Kumagai and Oba (KO) \cite{Kumagai-2014} where the correction energy is be expressed as:
\begin{equation}
\label{eq:corre2}
E_{corr} = -E_{lat} + q \Delta \phi
\end{equation}

where $E_{lat}$ is the Madelung energy of $\rho_d$ embedded in the host material and the jellium background when periodic boundary conditions are present and $\Delta \phi$ is an alignment term for the electrostatic potential. 
The scheme assumes that the defect-induced charge density is spherical and introduces a model charge density from which the terms are calculated. In the FNV scheme this is generally taken as a linear combination of a Gaussian and an exponential functions.
Using such a model for $\rho_d$ gives quite some flexibility for model the defect-induced charge density and allows to calculate analytically the Madelung energy for isotropic systems but analytic calculations for anisotropic systems are not possible. 
On the other hand, KO showed the loss of flexibility in using a point-charge (PC) model is generally small for most system and a PC model allows the analytic calculation of the Madelung energy in anisotropic systems \cite{Fischerauer-1997,Rurali_NL09}. As a matter of fact, the KO approach  has been successfully applied to a wide range of materials, showing that, once the defect-induced charge density is well localized within the supercell, a convergent defect formation energy can be obtained  using supercells of moderate size \cite{Kumagai-2014}.

The potential-alignment term, $\Delta \phi$, is obtained by comparing the electrostatic potential far from the defect position of the pristine and the defect containing supercells with the one of the model charge density. It can be decomposed into a sum of different terms \cite{Freysoldt-2009,Komsa_PRB12,Kumagai-2014}:

\begin{equation}
\label{eq:alignment_meaning}
\Delta \phi = V_{d, q} - V_{q/b} \equiv -\Delta V_{d, q/b},
\end{equation}
where $V_{d, q}$ is the electrostatic potential produced by the model charge distribution and $V_{q/b} = V_{\mathrm{defect}, q} - V_{\mathrm{bulk}}$ is the defect-induced potential as calculated in the first-principles simulation.
FNV perform the comparison using planar averages\cite{Freysoldt-2009}, whereas KO compare the atomic-site potentials in a region far from the defect.\cite{Kumagai-2014}. Provided large enough supercells, which allow for a sufficient sampling of the atomic-site potentials,  the latter method has been shown to provide better convergence, especially in the case of ionic materials where defect-induced atomic relaxations are large \cite{Kumagai-2014}.

\subsection{\label{subsec:concs} Equilibrium defect and carrier concentrations in the dilute limit}
Equilibrium defect and carrier concentrations can be calculated in the dilute limit once the relevant defect formation energies have been obtained.
In the dilute limit the defects do not interact and thus the energy required for forming $n$ defects of type $D$ in charge state $q$
is given by: $\Delta E_d(nD^{(q)}) = n\Delta E_d(D^{(q)})$.
A necessary condition for the thermodynamic equilibrium is that the system grand potential, $\Phi$, is at an extremum:
\begin{equation}
\label{eq:grand_pot}
\Phi(n D^{(q)}) = \Phi(\mathrm{bulk}) + \Delta E_d(nD^{(q)})  + T S_{conf}(n)
\end{equation}
where $S_{conf}$ is the contribution of the configurational entropy to the grand potential of the defect-laden system.
Assume that there are $g_{D^{(q)}}$ possible configurations in which $D^{(q)}$ has the same $\Delta E_d(D^{(q)})$.
Let $N$ be the number of unit cells forming the system and  $\gamma_{D^{(q)}}$ be the number of  equivalent sites in the unit cell that the defect can occupy, then (for $n \ll N\gamma_{D^{(q)}}$) the number of possible ways to place $n$ non-interacting defects on $N\gamma_{D^{(q)}}$ sites is:
\begin{equation}
\label{eq:dofs}
\Omega_{D^{(q)}} = g_{D^{(q)}} \binom{N\gamma_{D^{(q)}}}{n}
\end{equation}
yielding the configurational entropy $S_{conf} =  k_B \ln \Omega_{D^{(q)}}$. Finding the extremum of equation~\eqref{eq:grand_pot} with respect to $n$, using Stirling's approximation for $S_{conf}$, gives the equilibrium defect concentration:
\begin{equation}
\label{eq:conc}
c_{D^{(q)}} = \frac{n}{N} = \frac{\gamma_{D^{(q)}} g_{D^{(q)}}}{\exp{\left(\frac{\Delta E_d(D^{(q)})}{k_B T}\right)} + g_{D^{(q)}}}
\end{equation}
which is usually approximated by the limit value for $\Delta E_d({D^{(q)}}) \gg k_B T$ as:
\begin{equation}
c_{D^{(q)}} = g_{D^{(q)}} \gamma_{D^{(q)}} \exp{\left(-\frac{\Delta E_d({D^{(q)}})}{k_B T}\right)}
\end{equation}
In case of more than one type of defect in the crystal, the equilibrium concentration of each defect is still given by formula \eqref{eq:conc}, assuming the dilute-limit holds.

Equilibrium formation energies depend also on the electron chemical potential $\mu_e$ which needs to be evaluated before computing equilibrium concentrations. The equilibrium value of $\mu_e$ is fixed by the condition  that any actual solid will be characterized by a null net charge at equilibrium. \texttt{Spinney} computes $\mu_e$ by finding the roots of the equation describing the charge-neutrality condition:

\begin{equation}
\label{eq:charge_neutrality}
\sum qc_{D^{(q)}}(\mu_e) + p_0(\mu_e)  - n_0(\mu_e)  = 0,
\end{equation}
where $n_0$ is the concentration of free electrons:

\begin{equation}
\label{eq:el}
n_0 = \int_{\epsilon_C}^\infty \frac{n(\epsilon) }{e^{(\epsilon - \mu_e)/k_B T} + 1} d\epsilon,
\end{equation}
and $p_0$ the concentration of free holes:
\begin{equation}
\label{eq:holes}
p_0 = \int_{-\infty}^{\epsilon_V} \frac{n(\epsilon) }{e^{( \mu_e - \epsilon)/k_B T} + 1} d\epsilon.
\end{equation}
$n(\epsilon)$ is the density of states and $\epsilon_V, \epsilon_C$ are the eigenvalues of the valence band maximum and conduction band minimum, respectively.
Once the roots of equation \eqref{eq:charge_neutrality} have been found, equilibrium defect concentrations are obtained using equation \eqref{eq:conc} and carrier concentrations using equations \eqref{eq:el} and \eqref{eq:holes}.

Often dopants are introduced in the material in conditions which are far from the thermodynamic equilibrium assumed in the previous discussion, however, the thermodynamic formalism is still generally used to assess the properties of doped materials. If there is any indication that actual doping concentrations will noticeably differ from those predicted by thermodynamic equilibrium, \texttt{Spinney} allows to specify an effective doping concentrations $N_{\mathrm{eff}}$ quantifying the amount of ionized dopant species. The equilibrium electron chemical potential is then obtained by a modified version of equation \eqref{eq:charge_neutrality}:
\begin{equation}
\label{eq:charge_neutrality2}   
\sum qc_{D^{(q)}}(\mu_e) + p_0(\mu_e)  - n_0(\mu_e)  = N_{\mathrm{eff}}
\end{equation}

\section{Implementation and Examples}
\label{sec:examples}
\subsection{\label{subsec:general} General implementation features}

\begin{figure}
\centering
\includegraphics[width=.8\textwidth]{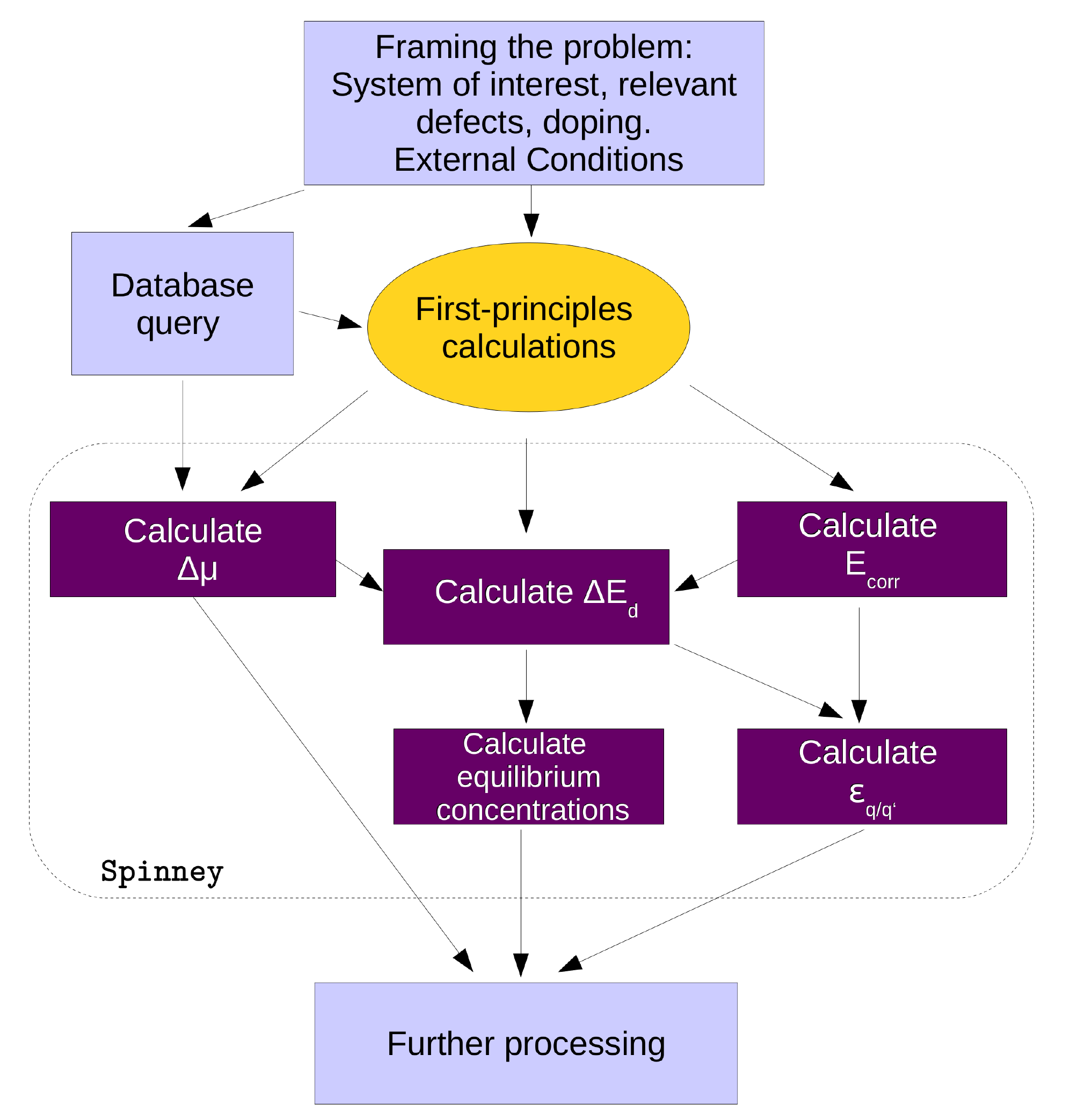}
\caption{\label{fig:workflow} Typical workflow for first-principles calculations of point defects. The properties which the \texttt{Spinney} code can calculate are included within the dashed box.}
\end{figure}
Figure~\ref{fig:workflow} illustrates the typical workflow for first-principles point defect calculations in solids. According to the problem of interest, one selects the most important native defects, eventual doping and the environmental conditions most pertinent for the applications of the material. Different environmental conditions entail different thermodynamic limits for the chemical potentials of the atomic species forming the system. In order to evaluate these limits, competing phases to the system under investigation must be considered. Multiple online repositories offer large databases of chemical compounds and the corresponding \emph{ab-initio} calculated electronic energy  and can be used for the identification of the most relevant phases. The most time-consuming part of the whole process consists in performing the first-principles calculations on the defect-containing supercells. Once the relevant calculations have been completed some post-processing of the first-principles data is required to obtain many of the energetic properties of the defect-laden system. At this stage \texttt{Spinney} comes into play: first-principles data are collected and fed to the appropriate routines which will output the quantities necessary for calculating defect formation energies, thermodynamic charge transition levels and equilibrium defects and carriers concentrations. The modular design of the code allows to calculate each of the defect properties independently. These results can then be used as input data for additional steps in the pipeline.

\texttt{Spinney} is written entirely in Python 3 and a rapid execution speed is achieved by extensive exploitation of \texttt{NumPy} arrays and the supported vectorized operations \cite{numpy}.
As mentioned in the introduction, the basic routines that allow for calculating the defect properties illustrated in Figure \ref{fig:workflow} accept built-in Python's data structures or \texttt{NumPy} arrays as input data. This allows the users of general first-principles codes to parse the data obtained in the calculations, format them and feed the result to the appropriate routine in order to calculate the desired property.  At the same time \texttt{Spinney} implements higher-level routines which can automatically parse the native output files of the popular DFT codes \texttt{WIEN2k} \cite{wien2k,blaha_JCP20} and \texttt{VASP} \cite{Kresse-1993, Kresse-1996}. For \texttt{VASP}, generally the \texttt{OUTCAR} and \texttt{vasprun.xml} files of the pristine and defect containing supercells are sufficient. The FNV alignment scheme furthermore requires the appropriate \texttt{LOCPOT} files with the calculated electrostatic potentials. The size of these files can be considerable, adding noticeable overhead  which makes the correction scheme often I/O bound. For \texttt{WIEN2k}, the files which are generally required are the \texttt{case.struct} and \texttt{case.scf}. The KO alignment scheme also requires the \texttt{case.vcoul} files. Additional information, which is brief, easily accessible and does not require \emph{ad hoc} parsers, such as defect positions, values of the dielectric tensor and valence band maximum, must be provided by the user.
The main output of \texttt{Spinney}'s basic routines are either scalars or tabular data which can be accessed also as \texttt{panda} dataframes \cite{pandas}, allowing  for seamless construction of databases for point defect in solids. Defect properties are also summarized by multiple plots enabled by the \texttt{Matplotlib} library \cite{Hunter-2007}.

\subsection{\label{subsec:exchem} Finding chemical potentials thermodynamic limits}
Taking Nb-doped TiO$_2$ anatase as an example, we will illustrate how to investigate the limit values for atomic chemical potentials using  \texttt{Spinney}. In particular, the \texttt{Spinney}'s module used for this analysis is \texttt{thermodynamics.chempots}.

The Ti-O system has a complex chemistry and many phases are know to exist. Niobium doping of anatase is a promising method for obtaining alternative transparent-conducting oxides \cite{Furubayashi-2005} and it is also employed for improving the photocatalytic properties of the material \cite{Su-2015}. The inequalities and equality constraints in equations~\eqref{eq:atchempots} define the feasible region the three atomic chemical potentials. In order to predict this region, we used the Materials Project's online repository \cite{Jain-2013}, from which we obtained the first-principles energies of more than 100 compounds in the Ti-O-Nb system. Formation energies for all these compounds where then calculated using as reference state the HCP structure for Ti, the BCC one for Nb and the triplet molecular state for O$_2$. As discussed in Ref.~\cite{Arrigoni_JCP20} the oxide materials formation energies were calculated including the term correcting for the binding energy of the O$_2$ molecule proposed in Ref.~\cite{Wang-2006}. 

\begin{table}
\centering
\begin{tabular}{l l c c c} 
 Case & & $ \Delta \mu_\text{Ti}$ (eV) & $ \Delta \mu_\text{O}$ (eV)  & $ \Delta \mu_\text{Nb}$ (eV) \\ 
 \hline
 \hline
\multirow{2}{3em}{1.} & min & -10.33 & -4.46 & $-\infty$  \\ 
 & max & -1.65 & -0.12 & 0 \\
 \hline
\multirow{2}{3em}{2.}  & min & -10.33 & -0.12 & $-\infty$ \\
 & max & -10.33 & -0.12 & -10.35 \\
 \hline
 \multirow{2}{3em}{3.}  & min & -1.78 & -4.46 & 0 \\
 & max & -1.65 & -4.39 & 0 \\
 \hline
\end{tabular}
\caption{Minimum and maximum values of $\Delta \mu$ calculated for the Ti-O-Nb system. The first row (1.) takes into account the whole feasible region. The second row (2.) considers the intersection of the feasible region with the $\Delta \mu_{\text O} = \Delta \mu_{\text O}^\mathrm{max}$ plane. The third row (3.) considers the intersection of the feasible region with the $\Delta \mu_\text{Nb} = 0$ plane. }
\label{table:chempots}
\end{table}

\begin{figure}
\subfloat[\label{fig:nbtichem}]{{\includegraphics[width=0.45\textwidth]{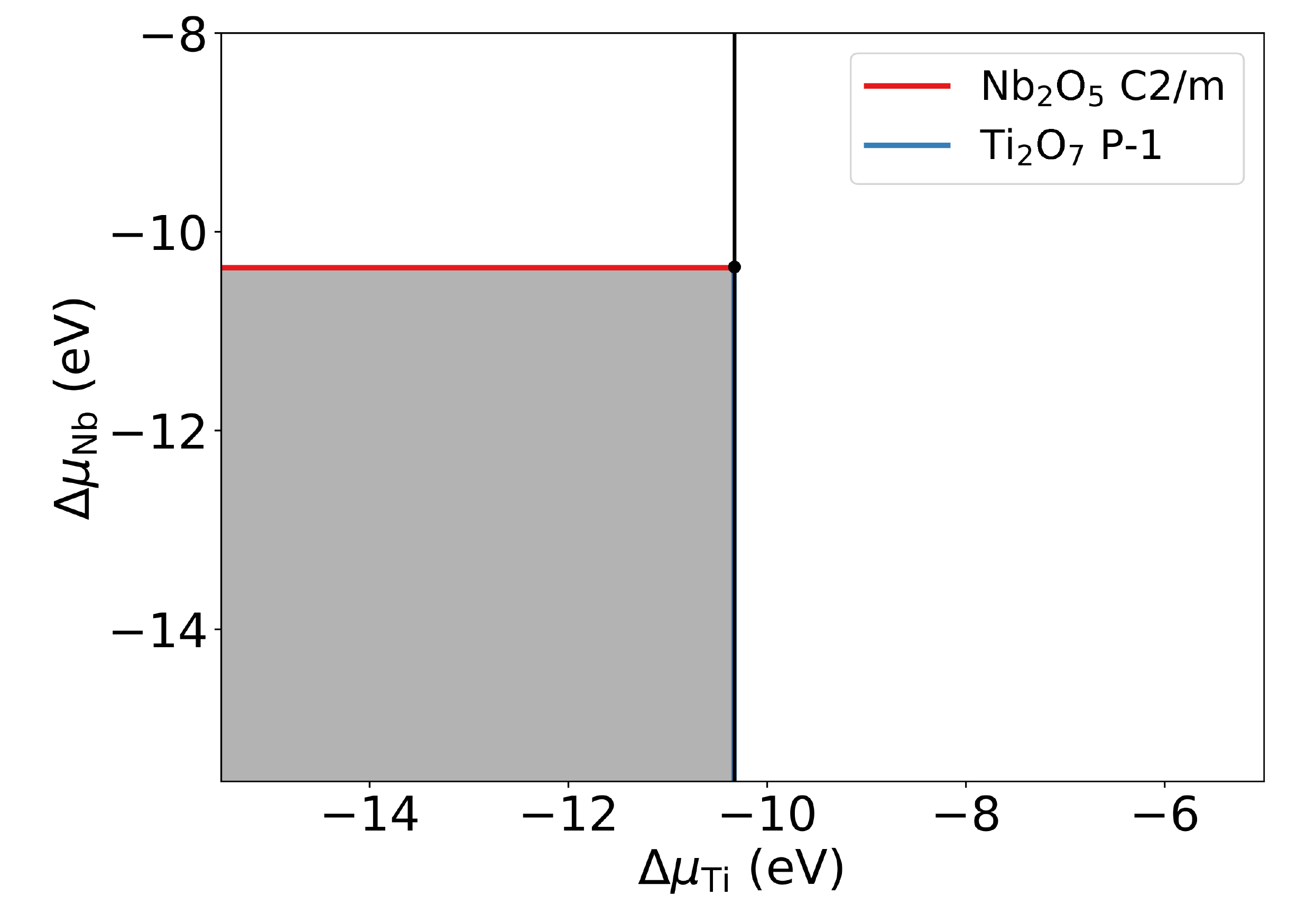} }} \hfill
\subfloat[\label{fig:tiochem}]{{\includegraphics[width=0.45\textwidth]{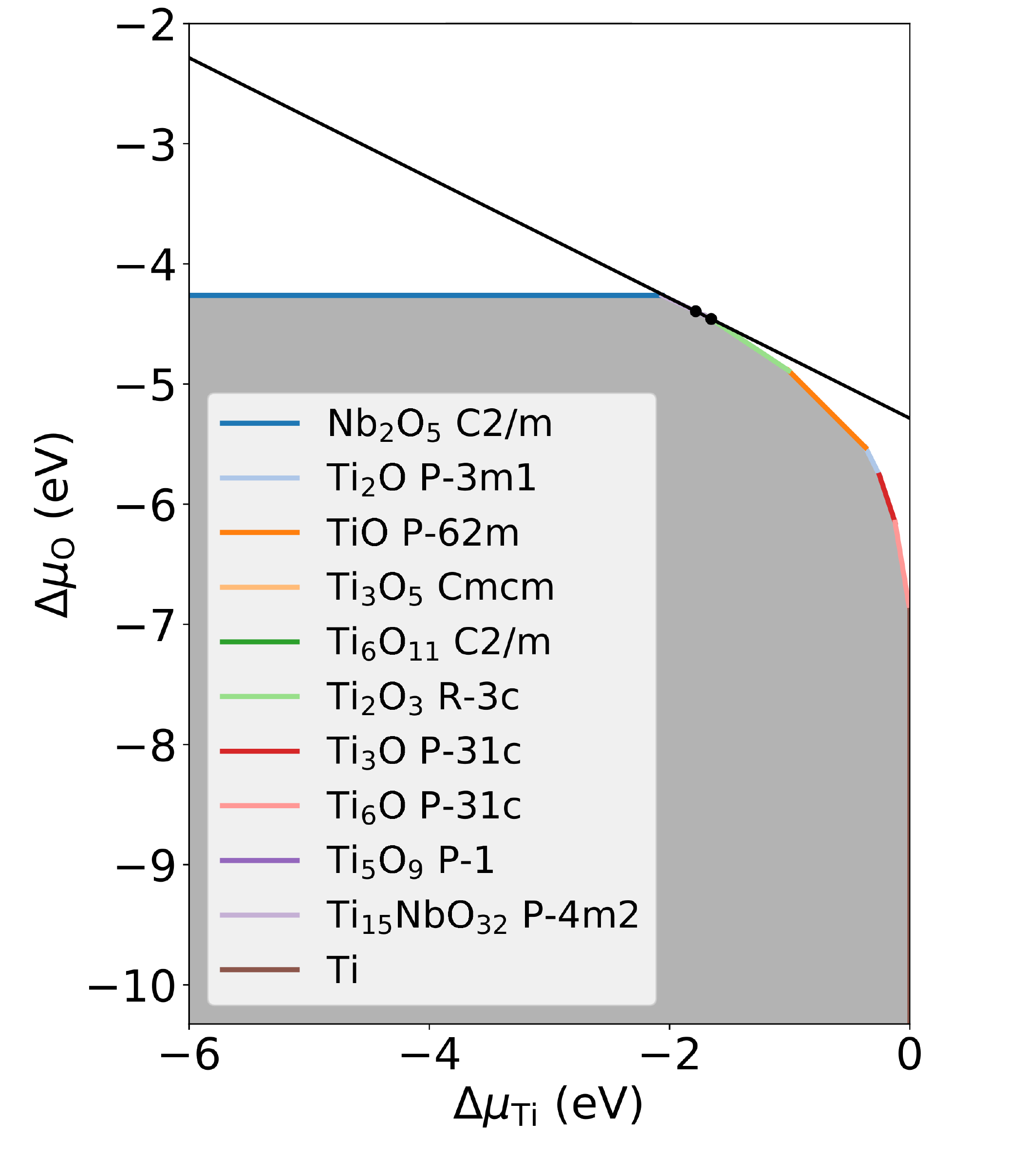}}}
\caption{\label{fig:chempots} Intersection of the feasible region of anatase TiO$_2$:Nb with: (a) the $\Delta \mu_\text{O} =  \Delta \mu_\text{O}^\mathrm{max} = -0.12$ eV plane; (b) the $\Delta \mu_\text{Nb} =  0$ eV plane. The gray area represents the inequalities constraints and the bold black line the equality constraint (see equations~\eqref{eq:equal_bound}). }
\end{figure}

Often defect properties, such as formation energies, are reported considering only the extreme values of the atomic chemical potentials, for which the system is in equilibrium with other phases. For example, assume one is interested in the O-rich limit and desire to find out the possible values that the chemical potentials  $\Delta \mu_\text{Ti}$ and $\Delta \mu_\text{Nb}$ can achieve. Figure~\ref{fig:nbtichem} shows the intersection of the feasible region with the $\Delta \mu_\text{O} = \Delta \mu_\text{O}^\mathrm{max} $ plane, where $ \Delta \mu_\text{O}^\mathrm{max}$, which is reported in Table~\ref{table:chempots},  represents the maximum value that  $\Delta \mu_\text{O}$ can achieve within the feasible region described by equations~\eqref{eq:atchempots}. As Figure~\ref{fig:nbtichem} and Table~\ref{table:chempots} show, thermodynamic equilibrium requires that the maximum value  $\Delta \mu_\text{Nb}$ can achieve is $-10.35$~eV in the O-rich limit. For higher values, Nb-doping would cause the segregation of $\mathrm{Nb_2O_5}$ in the $C2/m$ space group. Once $\Delta \mu_\text{O}$ has been fixed, $\Delta \mu_\text{Ti}$ is fixed as well due to the equality constraint.
Figure~\ref{fig:tiochem} shows another extreme condition where the feasible region intersects the $\Delta \mu_\text{Nb} = 0 $ plane. In this Nb-rich limit, $\Delta \mu_\text{Ti}$ and $\Delta \mu_\text{O}$ can vary over a narrow range, quantified in Table~\ref{table:chempots}, along the line describing the equality constraint. Outside this range, Ti$_2$O and Ti$_3$O$_5$ would start to precipitate.

\subsection{\label{subsec:fse} Corrections for electrostatic finite-size effects}

In this section we take as an illustrative example two intrinsic point defects in two wide-band-gap materials: the B vacancy in the charge state -3, $\square_\mathrm{B}^{-3}$, in cubic BN and the oxygen vacancy in the charge state +2, $\square_\mathrm{O}^{+2}$, in ZnO. 
\texttt{Spinney}'s modules for calculating electrostatic finite-size effects corrections employing the KO and FNV schemes are located in \texttt{spinney.defects}.

\begin{figure}
\subfloat[\label{fig:V_B}]{{\includegraphics[width=0.5\textwidth]{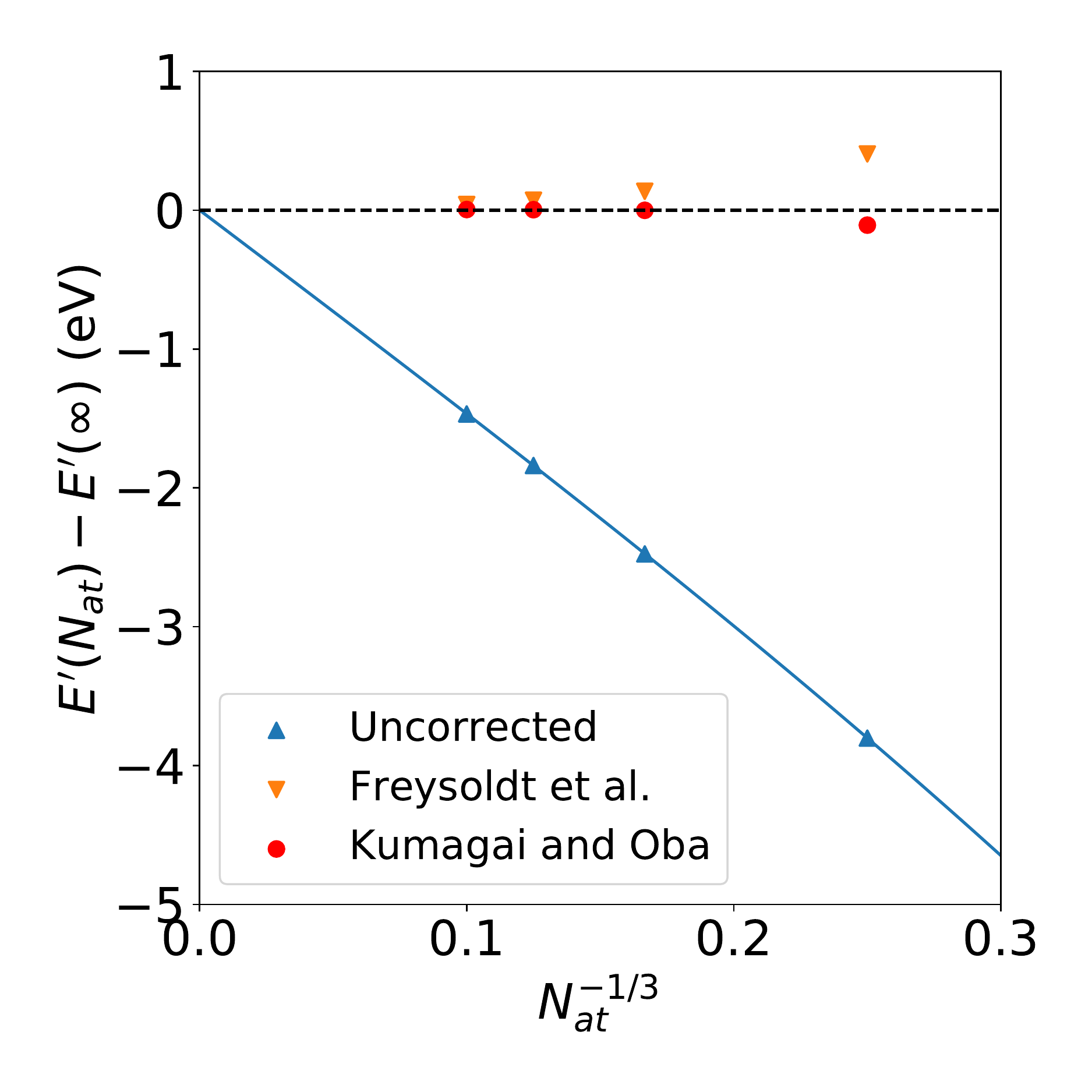}}} 
\subfloat[\label{fig:V_O}]{{\includegraphics[width=0.5\textwidth]{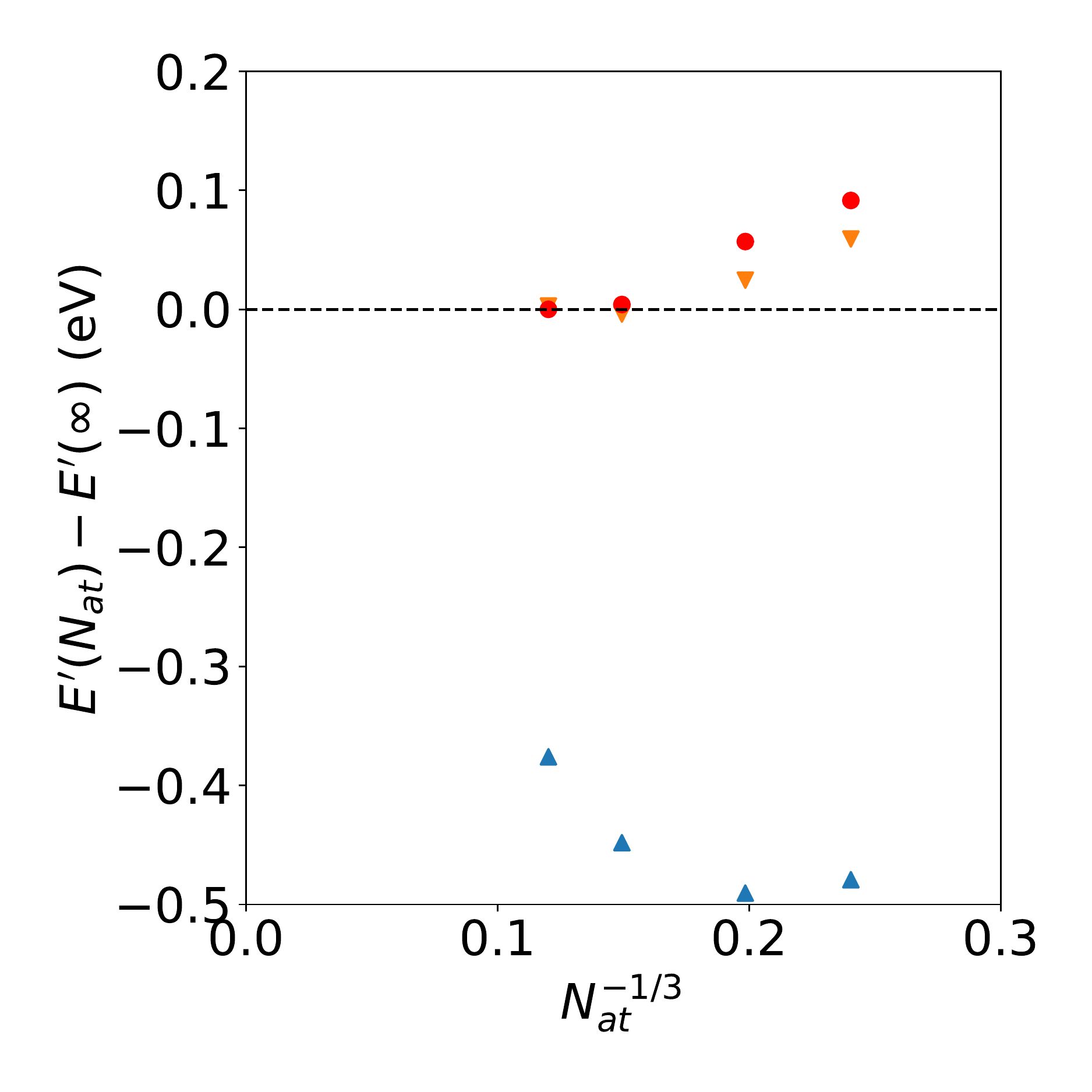}}}
\caption{\label{fig:eleccomp} Relative defect formation energy as a function of the supercell size. (a) $\square_\mathrm{B}^{-3}$ in c-BN; (b) $\square_\mathrm{O}^{+2}$ in ZnO. The zero of the energy scale represents the convergent value of the defect formation energy. For $\square_\mathrm{B}^{-3}$ in c-BN  the uncorrected defect formation energies are fitted  as: $E'(N_{at}) =   E'(\infty) + aN_{at}^{-1/3} + bN_{at}^{-1} $. In the scheme of Freysoldt \emph{et al.} we used a simple Gaussian charge density $N\exp(-r^2)$ where $N$ is the normalization and $r$ is given in \AA.}
\end{figure}

Figure~\ref{fig:eleccomp} compares the calculated defect formation energy as a function of the supercell size for  $\square_\mathrm{B}^{-3}$ in cubic BN (left-hand side) and for $\square_\mathrm{O}^{+2}$ in ZnO (right-hand side) using the two correction schemes for electrostatic effects implemented in \texttt{Spinney}. It can be observed that both correction schemes predict the same value of the defect formation energy for large enough supercells. In particular, Figure \ref{fig:V_B} shows that for an isotropic system the predicted value of the defect formation energy, after applying the correction term, converges to the limit value for an infinite large supercell $E'(\infty)$, described by the heuristic equation: $E'(N_{at}) =  E'(\infty) + aN_{at}^{-1/3} + bN_{at}^{-1} $, where $E'(N_{at})$ represents the \emph{uncorrected} energy of a supercell containing $N_{at}$ atoms \cite{Castleton-2006}.\

\begin{table}
\centering
\begin{tabular}{l l |c c c } 
  \hline
 System &  Supercell &  $-E_{lat}$ (eV) & $q\Delta \phi_{KO}$ (eV)  &  $q\Delta \phi_{FNV}$ (eV) \\ 
 \hline
  & $2\times 2 \times 2$ &  3.67 & 0.03  & 0.53 \\
  $\square_\mathrm{B}^{-3}$ & $3\times 3 \times 3$ & 2.44 & 0.03 & 0.17 \\
 c-BN& $4\times 4 \times 4$ & 1.83 & 0.09 &  0.08 \\
  & $5\times 5 \times 5$ &  1.47 & 0.01 &  0.04 \\
  \hline
  & $3\times 3 \times 2$ &  0.80 & -0.23 &  -0.25 \\
 $\square_\mathrm{O}^{+2}$ & $4\times 4 \times 2$ & 0.66 & -0.11  & -0.14 \\
 ZnO & $5\times 5 \times 3$ &  0.50 & -0.05 &  -0.05 \\
 & $6\times 6 \times 4$ &  0.40 & -0.02 & -0.02 \\
 \hline
\end{tabular}
\caption{Contribution to the correction energy for electrostatic finite-size effects calculated with the method of Kumagai and Oba and with the one of Freysoldt \emph{et al.} for two type of point defects. For c-BN, a supercell expansion of the conventional cubic cell is used so that the $2\times2\times2$ cell contains 64 atoms. The $3\times3\times2$ ZnO cell contains 108 atoms. For $\square_\mathrm{B}^{-3}$ the B-rich, $\Delta \mu_\mathrm{B} = 0$, and for $\square_\mathrm{O}^{+2}$ the O-rich, $\Delta \mu_\mathrm{O} = 0$, limits were considered. All first principles calculations were performed using the Perdew, Burke and Ernzerhof (PBE)  \cite{Perdew-1996} functional and the projector augmented-wave (PAW) method \cite{Bloech-1994}. For these two systems, $E_{lat}$ calculated by the two methods differs in all case by at most 0.01 eV and is therefore reported only once.}
\label{table:compcorr}
\end{table}

\begin{figure}
\centering
\includegraphics[width=0.5\textwidth]{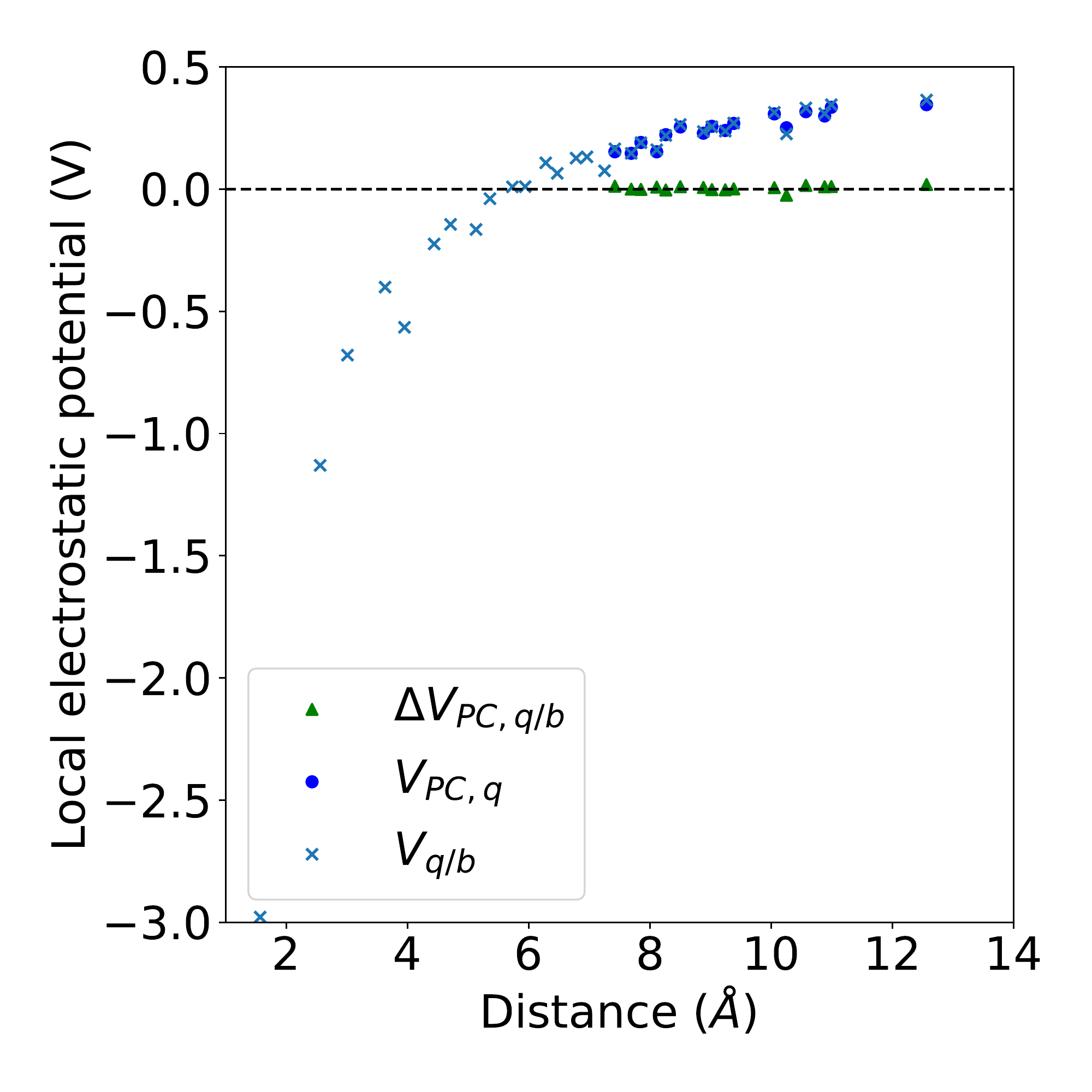}
\caption{\label{fig:align} Local electrostatic potentials used to compute the potential alignment term, $\Delta \phi$, in the scheme of Kumagai and Oba, plotted with respect to the distance from $\square_\mathrm{B}^{-3}$ in c-BN using a $4 \times 4 \times 4$ supercell expansion. The atomic site potentials are used: $V_{q/b}$ is the difference between the pristine and defect containing supercell potentials obtained from the first-principles calculations, $V_{PC, \, q}$ is the potential generated by the point charge model and $\Delta V_{PC, \, q/b}$ is the difference between these two potentials. The latter term is used to calculate $\Delta \phi$ and its value is well converged in the sampling region far from the defect. The dashed horizontal line indicate the zero of the potential. 
}
\end{figure}

Table~\ref{table:compcorr} compares the two terms entering equation \eqref{eq:corre2} calculated in the two correction schemes. 
 The method of KO generally allows for a more robust potential alignment procedure since it was found in Ref.~\cite{Kumagai-2014} that the atomic-site potentials are able to converge much faster far from the defect in ionic materials; while this is not generally the case for the plane-averaged electrostatic potential. 
Potential-vs-distance plots like those in Figure~\ref{fig:align} or Figure~1 of Ref.~\cite{Kumagai-2014}  represent therefore a valuable tool for assessing the accuracy of the correction scheme and can be readily obtained with \texttt{Spinney}.

We finally compared the defect formation energies of $\square_\mathrm{B}^{-3}$ in c-BN calculated obtained with VASP and WIEN2k. The computed values are in very good agreement between the two codes. The difference in $\Delta E_d$ is about 0.2 to 0.3~eV for the $3\times3\times3$ and $4\times4\times4$ cells which amounts to about $2 \%$.

\subsection{\label{subsec:ctl} Thermodynamic charge-transition levels}
In these last two sections, Mg-doped GaN is taken as an illustrative example. GaN is a very important semiconductor material which finds applications in photodetectors, light emitting diodes (LEDs) in the blue and ultraviolet region, laser diodes and bipolar transistors \cite{Ambacher-1998, Morkoc-1994, Johnson-2002}. The material shows an intrinsic n-type conductivity. On the other hand, for improving the properties of GaN-based electronic devices such as LEDs and lasers the synthesis of p-type GaN is desirable. Obtaining p-type GaN is generally challenging but Mg is one of the most successful dopants used for this purpose \cite{Amano-1989, Akasaki-1991}.
 
This section reports thermodynamic charge transition levels for GaN:Mg. The next section, section \ref{subsec:fconcs}, will use these results for calculating defects and carriers equilibrium concentrations. The employed \texttt{Spinney}'s modules are located in \texttt{spinney.defects}. All calculation were performed using the VASP code and PAW pseudopotentials with the PBE exchange-correlation functional. A supercell containing 96 atoms was used and corrections for electrostatic finite-size effects for charged defects were included using the KO scheme. All intrinsic point defects (Ga and N vacancies, Ga and N intersitials, Ga$_\mathrm{N}$ and N$_\mathrm{Ga}$  antisites) and the Mg$_\mathrm{Ga}$ substitutional impurity were considered.

\begin{figure}
\centering
\subfloat[\label{fig:Garich}]{{\includegraphics[width=0.5\textwidth]{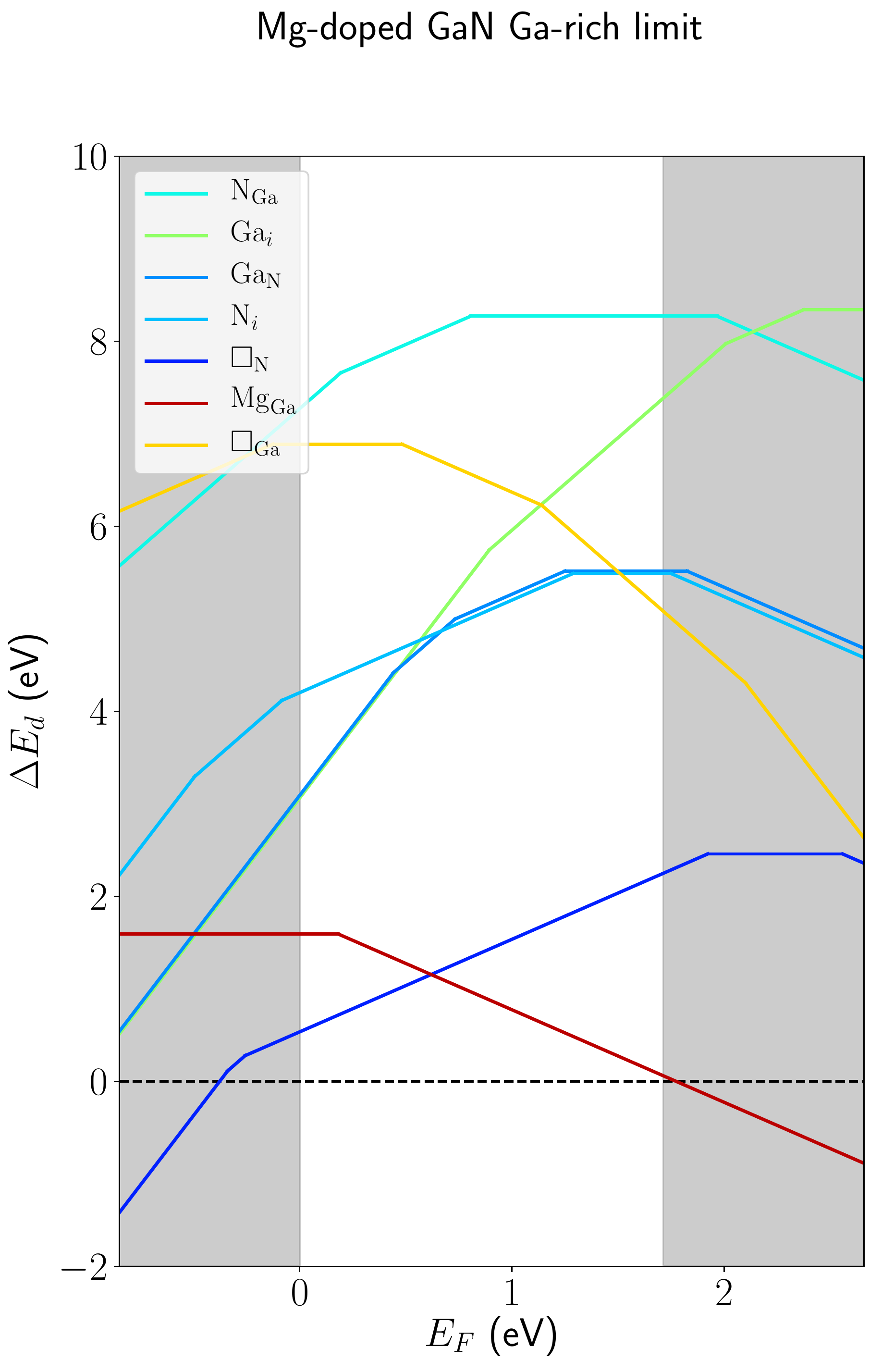}}} 
\subfloat[\label{fig:Nrich}]{{\includegraphics[width=0.5\textwidth]{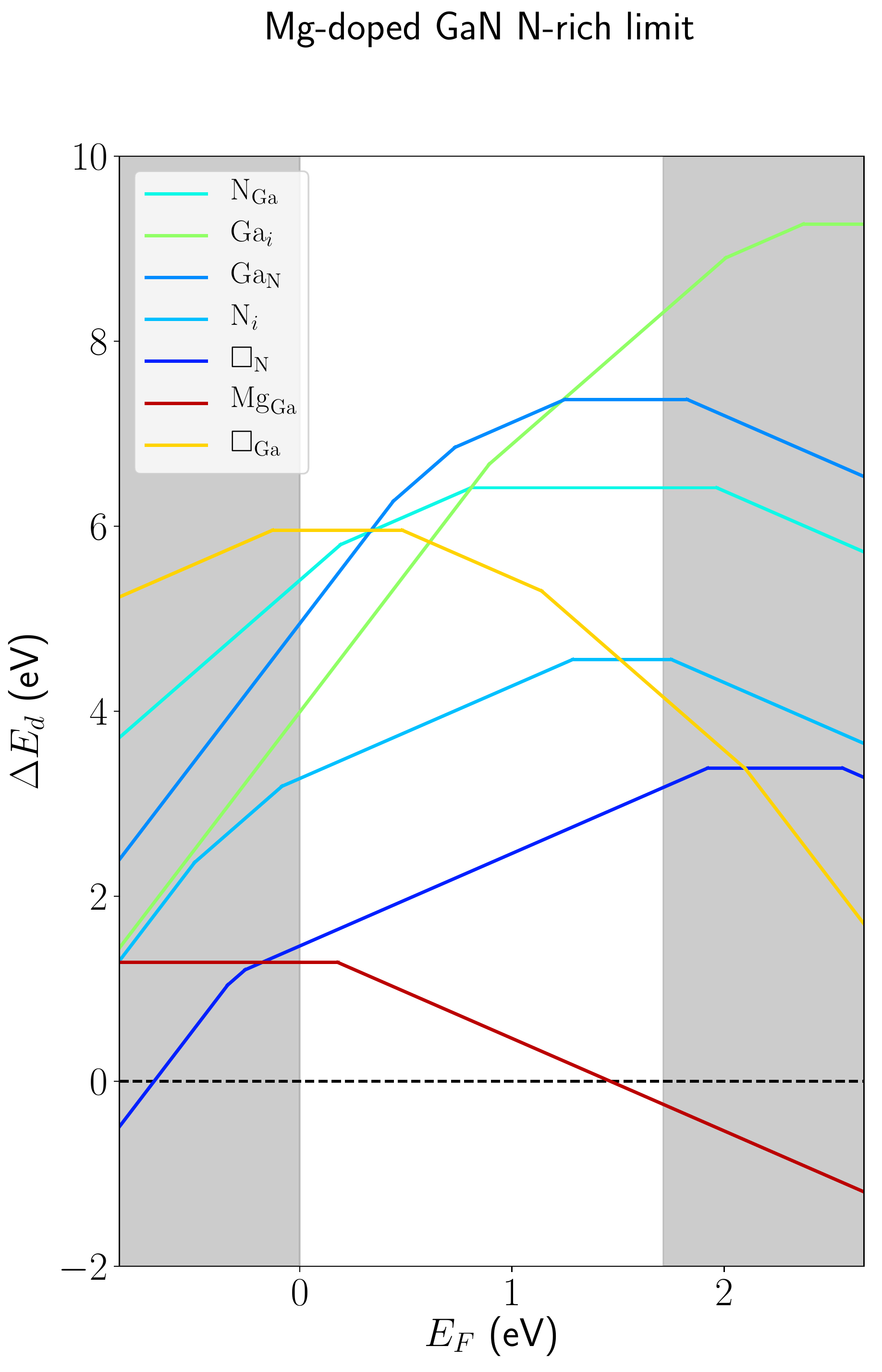}}}
\caption{\label{fig:ctl} Defect formation energies as a function of the Fermi level. The white are represents the fundamental band gap predicted by the PBE functional. This area is extended by the gray regions to the HSE fundamental gap, aligning the valence band maxima of the two functionals. (a) Ga-rich conditions, (b) N-rich conditions.}
\end{figure}

Figure~\ref{fig:ctl} shows the calculated $\Delta E_d$ as a function of the Fermi level, $E_F$, in Ga-rich ($\Delta \mu_\mathrm{Ga} = 0$ eV, left-hand side) and N-rich ($\Delta \mu_\mathrm{N} = 0$ eV, right-hand side) conditions. Using the same approach illustrated in section \ref{subsec:exchem}, we found that in both limits, the atomic chemical potential values are constrained by the formation of $\mathrm{Mg_3N_2}$. In the former limit, at equilibrium with $\mathrm{Mg_3N_2}$, $\Delta \mu_\text{N} = -0.93$ eV and $\Delta \mu_\text{Mg} = -0.67$ eV. In the latter limit, since $\mathrm{Mg_3N_2}$ is still the competing phase, $\Delta \mu_\text{Ga} = -0.93$ eV, while  $\Delta \mu_\text{Mg} = -1.29$ eV. These chemical potential values were used to calculate defect formation energies in Figure~\ref{fig:ctl}. 
The calculated charge transition levels for selected point defects in GaN:Mg are reported in Table~\ref{table:translev}.
As charge transition levels are defined for values of $E_F$ within the band gap, they are highly affected by the well-known DFT band-gap error. It has been shown that quantitative predictions of the charge transition levels in better agreement with more accurate functionals, such as hybrid ones, can be obtained by aligning the valence bad maximum predicted by local/semilocal functionals with the one of the more accurate functional \cite{Alkauskas-2008, Alkauskas-2011}. Figure \ref{fig:ctl} displays such alignment: the PBE band gap (white area) is extended (through the gray regions) so that the conduction band maxima of the PBE functional and the hybrid functional of Heyd, Scuseria and Ernzerhof (HSE) \cite{Heyd-2003} are aligned. The offset between the PBE and HSE valence band maxima has been taken  from Ref.~\cite{Lyons-2017}, which considers the same system and the same functionals. While,  after such an alignment has been performed, PBE charge transition levels predicted for $\square_\mathrm{N}$ agree  discretely well with HSE calculations (\emph{cf.} Ref.~\cite{Lyons-2017}), this is not the case for $\mathrm{Mg}_\mathrm{Ga}$ which is predicted by PBE to be a quite deep acceptor, while HSE predicts it to be much shallower \cite{Lyons-2012}. This feature is not surprising as the $\mathrm{Mg}_\mathrm{Ga}$ defect is an uncommon shallow acceptor, and HSE calculations have found that it is characterized by a  strongly localized hole on a neighboring N atom \cite{Lyons-2012}, which cannot correctly be described by PBE due to the well known self-interaction error.  

\begin{table}
\centering
\begin{tabular}{l r c } 
 Defect &  $q/q'$ & $\epsilon(q/q')$ (eV) \\ 
 \hline
 \multirow{4}{4em}{$\square_\mathrm{N}$} & 2/3 & 0.51 \\
 & 1/2 & 0.59  \\
 & 0/1 & 2.77  \\
  & -1/0 & 3.41 \\
  \hline
 \multirow{1}{4em}{$\mathrm{Mg}_\mathrm{Ga}$} & -1/0 & 1.03 \\
 \hline

\end{tabular}
\caption{Thermodynamic charge transition levels of the most relevant defects in GaN:Mg calculated using the PBE functional. The zero of the Fermi level is set to the top of the HSE valence band.}
\label{table:translev}
\end{table}

\subsection{\label{subsec:fconcs} Equilibrium defects and carriers concentrations}
Once defect formation energies have been computed for the defects of interest, equilibrium defect concentrations in the dilute limit can be calculated with the formalism presented in section \ref{subsec:concs}. Bulk GaN is usually growth at high temperatures and the thermodynamic conditions can be described as Ga-rich due to the very high nitrogen equilibrium pressure at these temperaratures \cite{Perlin-1995}. Intrinsic GaN  shows high electron concentrations in the range of 10$^{17}$-10$^{20}$ cm$^{-3}$ \cite{Perlin-1995, Kamler-2000} and there is  general consensus among experimental and theoretical studies that they arise from the ionization of nitrogen vacancies which are present at high concentration at the growth conditions (see Ref.~\cite{Perlin-1995} and references therein).

\begin{figure}
\centering
\includegraphics[width=.5\textwidth]{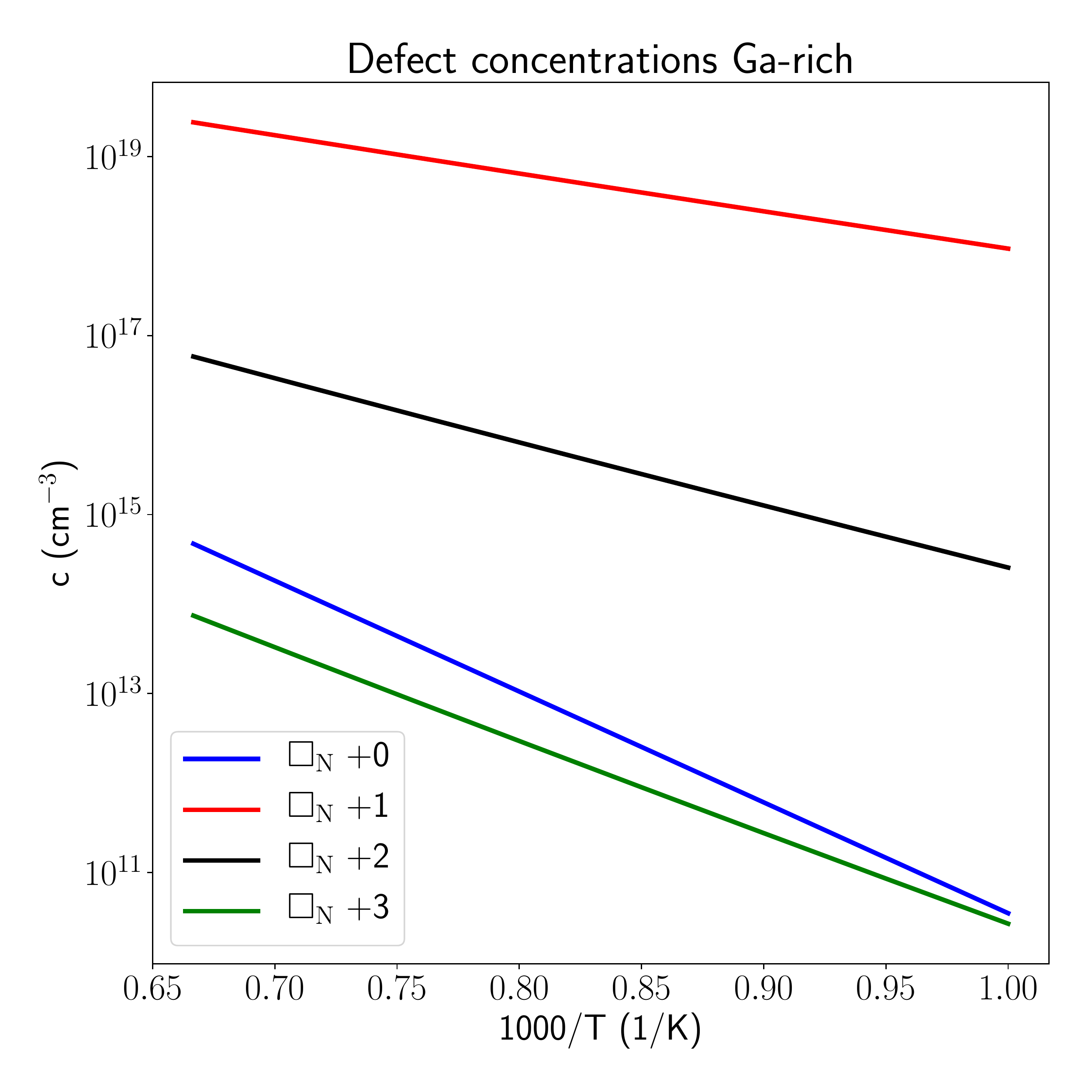}
\caption{\label{fig:v_conc} Equilibrium concentrations of $\square_\mathrm{N}$ in intrinsic GaN for high-temperature (1000-1500 K) and Ga-rich conditions.}
\end{figure}

Figure \ref{fig:v_conc} shows the equilibrium concentrations of the N vacancy in Ga-rich conditions calculated for intrinsic GaN considering a high-temperature range representing the experimental growth conditions of bulk samples. From the calculations, only $\square_\mathrm{N}$ in the illustrated charge states assumes concentrations larger than 10$^{10}$ cm$^{-3}$ in the considered temperature range. This indicates that the high concentrations of free electrons do indeed stem from the ionization of N vacancies and, in particular, from singly ionized donors, whose predicted concentration ranges from 10$^{18}$ to 2$\times$10$^{19}$ cm$^{-3}$ in the temperature range between 1000 and 1500 K.

\begin{figure}
\centering
\includegraphics[width=.5\textwidth]{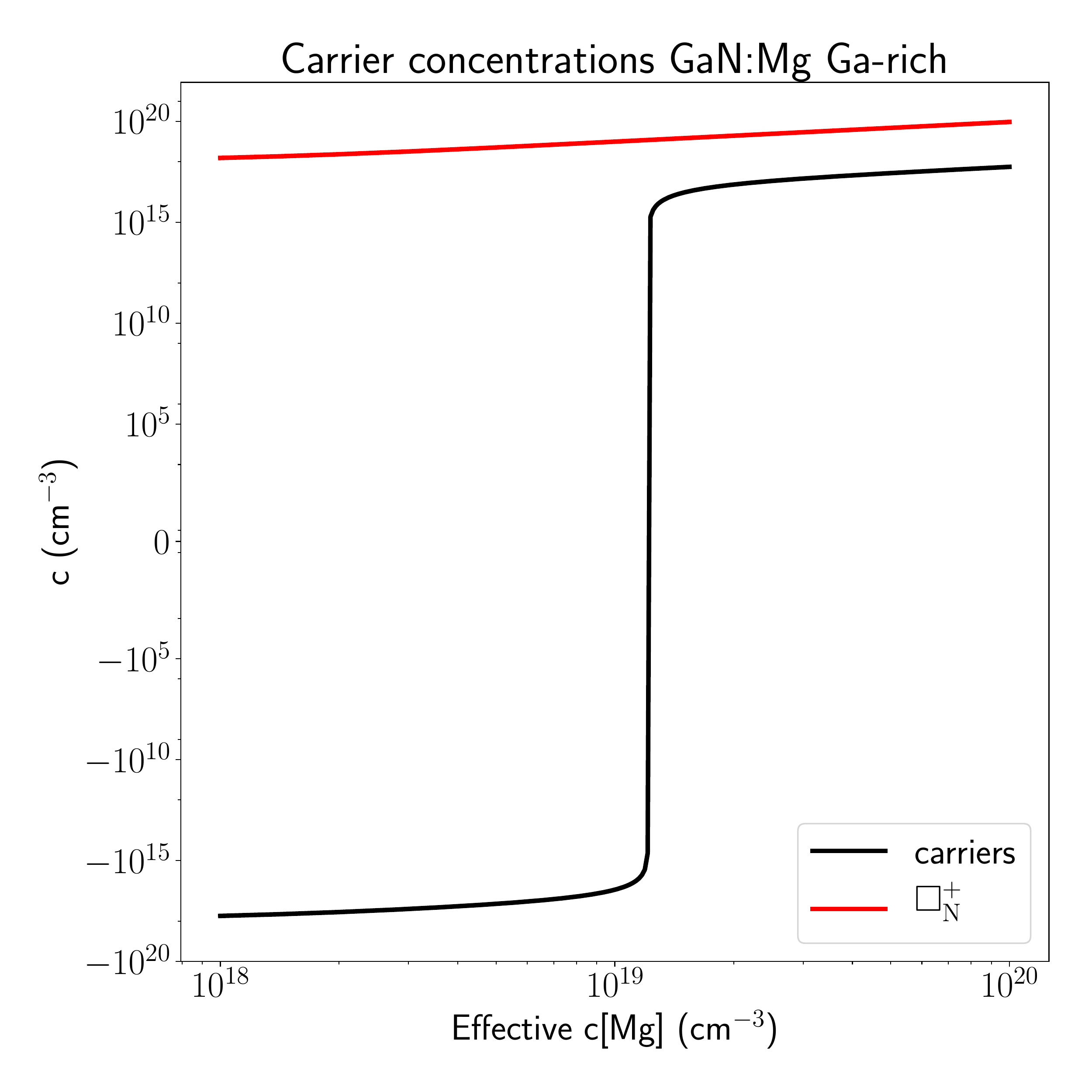}
\caption{\label{fig:carriers} Equilibrium concentrations of carriers and $\square_\mathrm{N}^+$ as a function of the effective concentration of Mg calculated for Ga-rich conditions and a growth temperature of 1000 K. For the carrier concentrations, negative values represent electrons  and positive ones holes.}
\end{figure}

Figure \ref{fig:carriers} shows the calculated carrier concentrations as a function of the effective concentration of Mg doping, which represents the amount of activated single-acceptor impurities at a given temperature (\emph{cf.} with equation \eqref{eq:charge_neutrality2}). 
Ga-rich conditions and  a growth temperature of 1000 K were considered. The concentration of the most relevant donor species, $\square_\mathrm{N}^{+}$, is also shown. From the picture it is clear that the acceptor doping is compensated by an increase of the intrinsic donor $\square_\mathrm{N}^{+}$ but hole concentrations larger than 10$^{16}$  cm$^{-3}$ can be obtained for an effective Mg doping larger than 10$^{19}$  cm$^{-3}$. This model predicts a monotone increase in the hole concentration as the amount of doping increases. In practice, the dilute limit theory breaks down for large dopant concentrations and segregation of Mg, with a decline of the p-type conductivity, has indeed been observed for large Mg concentrations (see  \cite{Paskov-2018} and references therein).

\section{Conclusion}
\label{sec:conclusion}
We have presented \texttt{Spinney} a Python 3 package for post-processing first principles calculations of point defects in semiconductors. Based on the theory of solid state solutions in the dilute limit, the package is able to calculate the most relevant energetic properties of point defects, including formation energies and thermodynamic charge transition levels, and applying state-of-the-art correction schemes for electrostatic finite-size effects. 
The package can be used as a Python module, making it easy to integrate with other computational frameworks.
In this contribution we have shown in detail how the package can be used to analyse and predict the properties of materials of technological relevance.

\section{Acknowledgements}
M.A would like to thank Peter Blaha for the fruitful discussions. 
The authors acknowledge support from the Austrian Science Funds (FWF) under project CODIS (FWF-I-3576-N36). We also thank the Vienna Scientific Cluster for providing the computational facilities (1523306: CODIS).

\bibliographystyle{elsarticle-num}

\end{document}